\documentclass{elsart}
\usepackage{epsfig}
\begin{document}
\begin{frontmatter}
\title{Precision Higgs physics at a $\gamma \gamma$ collider} 
\author[Rome]{Michael Melles}

\address[Rome]{Paul Scherrer Institute, Villigen, CH-5232, Switzerland.}

\begin{abstract}
The loop induced coupling of an intermediate mass Higgs boson 
to two photons is a sensitive and
unique measure for precision tests of physics beyond the
Standard Model.
In this work we
summarize recent results on the expected precision of the partial
$\Gamma ( H \rightarrow \gamma \gamma )$ width at the $\gamma \gamma$
option
of a future linear collider. 
Heavy particles do not decouple in general and differences between the SM
and
MSSM predictions or 2HD-models can differ in the percentile regime.
Large non-Sudakov DL corrections need to be resummed and consistency 
requirements demand the use of the Sterman-Weinberg jet definition
in order to avoid additional DL terms from three jet final states.
We find that the well understood background
process $\gamma \gamma \rightarrow q \overline{q}$ allows for a
${\mathcal O}$(2\%) 
determination of $\Gamma ( H \rightarrow \gamma \gamma )$ using
conservative collider parameters.
Recent improvements in the expected $\gamma \gamma$ luminosity suggest that
the precision for the diphoton partial Higgs width can be further
improved and is dominated by the
error in BR($H \rightarrow b \overline{b}$) from the $e^\pm$ mode, which is
presently estimated to be in the one percent regime.
\end{abstract}
\end{frontmatter}

\begin{center}
\begin{figure}
\centering
\epsfig{file=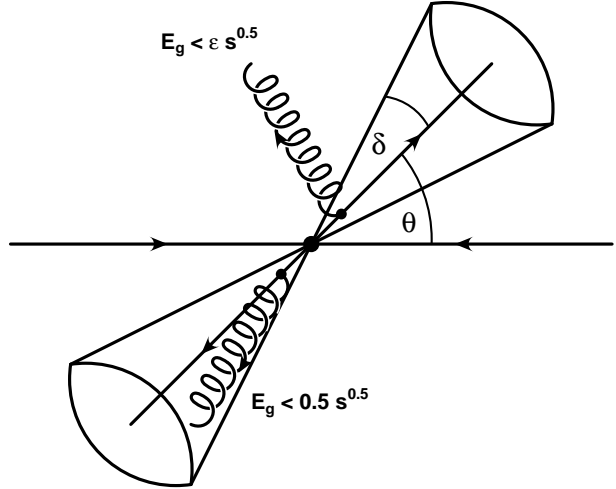,width=8cm}
\caption{The parameters of the Sterman-Weinberg two-jet definition used in this
work. Inside an angular cone of size $\delta$ arbitrary hard gluon bremsstrahlung
is included. Radiation outside this cone is only permitted if the gluon energy
is below a certain fraction ($\epsilon$) of the incident center of mass energy.
The thrust angle is denoted by $\theta$.}
\label{fig:jet}
\end{figure}
\end{center}

The partial diphoton Higgs width $\Gamma ( H \longrightarrow
\gamma \gamma)$, measured at the $\gamma \gamma$ Compton-backscattered option
of a future linear $e^\pm$ collider, is a very important physical 
quantity \cite{bbc}.
In Ref. \cite{ddhi}
it was found that the MSSM and SM predictions can differ in the percentile
regime for large masses of the pseudoscalar Higgs $m_A$, depending mainly
on the chargino-masses.
The SM with two Higgs doublets (2HDM) and all other Higgs particles heavy
differs by about 10\% \cite{gko}.
At the PLC one measures the product $\Gamma ( H \longrightarrow \gamma \gamma)
\times
BR ( H \longrightarrow b \overline{b})$ and it is assumed that the branching ratio
can be measured in the $e^\pm$ mode via $BR ( H \longrightarrow b \overline{b})=
\frac{[\sigma (ZH) \times BR ( H \longrightarrow b \overline{b})]}{
\sigma (ZH)}$ with a 1 \% accuracy \cite{br}.
It was recently demonstrated in Ref. \cite{msk,m} that using conservative assumptions
an accuracy of 2\% is feasible for the diphoton partial width 
at a PLC. 
There has been considerable progress in the theoretical understanding of the
BG to the intermediate mass Higgs boson decay into $b \overline{b}$ recently.
The Born cross section for the $J_z=0$ channel is suppressed by $\frac{m_q^2}{
s}$ relative to the $J_z=\pm2$ which means that by ensuring a high degree of
polarization of the incident photons
one can 
{\it simultaneously} enhance the signal and suppress the background.
QCD radiative corrections can remove this suppression, however, and large 
bremsstrahlung
and double logarithmic corrections need to be taken into account. 
In Ref. \cite{jt} the exact one loop corrections to $\gamma \gamma 
\longrightarrow q \overline{q}$ were calculated and the largest virtual correction
was contained in novel non-Sudakov double logarithms. For some choices of
the invariant mass cutoff $y_{cut}$ even a negative cross section was obtained
in this approximation. The authors of Ref. \cite{fkm} elucidated the physical
nature of the novel double logarithms and performed a two loop calculation
in the DL-approximation. The results restored positivity to the physical
cross section. In Ref. \cite{ms1}, three loop DL-results were presented which
revealed a factorization of Sudakov and non-Sudakov DL's and led to the
all orders resummation of all DL in form of a confluent hypergeometric function
$_2F_2$. The general form of the expression is $\sigma_{DL}=\sigma_{Born}(1+
{\mathcal F}_{DL}) \exp({\mathcal F}_{Sud})$.
In Ref. \cite{ms2} it was demonstrated that at least four loops on the
cross section level are required to achieve a converged DL result. 
At this
point the scale of the QCD-coupling is still unrestrained and differs by
more than a factor of two in-between the physical scales of the problem,
$m_q$ and $m_H$. This uncertainty was removed in Ref. \cite{ms3} by introducing
a running coupling $\alpha_s ({\bf l^2_\perp})$ into each loop integration, 
where $l_\perp$ denotes the perpendicular Sudakov loop momentum. 
The effect of the RG-improvement 
lead to
$\sigma^{RG}_{DL}=\sigma_{Born}(1+
{\mathcal F}^{RG}_{DL}) \exp({\mathcal F}^{RG}_{Sud})$.
The effective scale, defined simply as the one
used in the DL-approximation which gives a result close to the RG-improved
values, depends on $\epsilon$, however in general is rather much closer to
$m_q$ than $m_H$ \cite{ms3}.
On the signal side, the relevant radiative corrections have long been known
up to NNL order in the SM \cite{dsbz,dsz} and are summarized including the MSSM
predictions in Ref. \cite{s}. 
For our purposes the one loop corrections
to the diphoton partial width are sufficient as the QCD corrections
are small in the SM.
The important point to make here and also the novel feature in this analysis is
that the branching ratio BR ($H \longrightarrow b \overline{b}$) is corrected
by the same RG-improved resummed QCD Sudakov form as the continuum heavy quark background
\cite{msk}. This is necessary in order to employ the same two jet definition for
the final state. Since we use the renormalization group improved 
massive Sudakov form factor ${\mathcal F}^{RG}_{Sud}$ of Ref. \cite{ms3}, we prefer the
Sterman-Weinberg jet definition \cite{sw} schematically depicted in Fig. 
\ref{fig:jet}. 
This is also necessitated by the fact that for three jet-topologies new
DL corrections would enter which are not included in the background
resummation of Ref. \cite{ms1}.
We also use an all orders resummed running quark mass evaluated
at the Higgs mass for $\Gamma ( H \longrightarrow b \overline{b})$. For
the total Higgs width, we include the partial Higgs to $b \overline{b}, c
\overline{c}, \tau^+\tau^-, WW^*, ZZ^*$ and $gg$ decay widths with all relevant
radiative corrections.
\vspace{-1cm}
We begin with a few generic remarks concerning the uncertainties in our
predictions. The signal process $\gamma \gamma \longrightarrow H \longrightarrow
b \overline{b}$ is well understood and NNL calculations are available. The
theoretical error is thus negligible \cite{s}.
There are two contributions to
the background process
$\gamma \gamma \longrightarrow q \overline{q}$ which we neglect in this paper.
Firstly, the so-called resolved photon contribution
was found to be a small effect, e.g. \cite{jt}, especially since
we want to reconstruct the Higgs mass from the final two-jet measurements and
impose angular cuts in the forward region. In addition the good charm suppression
also helps to suppress the resolved photon effects as they give the
largest contribution.
The second contribution we do not consider here results from the final state 
configuration where
a soft quark is propagating down the beam pipe and the gluon and remaining
quark form two hard back-to back-jets \cite{bkos}. We neglect this contribution
here due to the expected excellent double b-tagging efficiency and the strong
restrictions on the
allowed acollinearity discussed below.
A good measure of the remaining theoretical uncertainty in the continuum background
is given by scanning it below and above the Higgs resonance. For precision extractions
of $\Gamma ( H \longrightarrow \gamma \gamma )$ the exact functional form for resonant
energies is still required, though.
In terms of possible systematic errors, the most obvious effect comes from the 
theoretical uncertainty
in the bottom mass determination. Recent QCD-sum rule analyses, however, reach below the 2\% level
for $\overline{m}_b(\overline{m}_b)=4.17 \pm 0.05$ \cite{hg} 
including the effect
of a massive charm \cite{m1}. 
For quantitative estimates of expected systematic experimental
errors it is clearly too early to speculate at this point. The philosophy adopted henceforth is that
we assume that they can be neglected at the 1\% level and concentrate purely on the statistical
error.
We focus here not on specific predictions for cross sections, but instead on the
expected statistical accuracy of the intermediate mass Higgs signal at a PLC.
As detailed in Refs. \cite{bbc}, due to the narrow Higgs width, the signal event
rate is proportional to $N_S \sim \left. \frac{dL_{\gamma \gamma}}{dw} 
\right|_{m_H}$, while the BG is proportional to $L_{\gamma \gamma}$. 
To quantify this, we take the design parameters of the proposed TESLA
linear collider  \cite{t,tp},
which correspond to an integrated peak
$\gamma \gamma$-luminosity\footnote{As reported by V.~Telnov at this meeting
\cite{tgg},
the luminosity can potentially be increased by
an order of magnitude. 
For Higgs energies it might be feasible to increase the luminosity
by a factor of 15 due to a
possible decrease of the horizontal beam emittance at the damping ring and
an increase of the repetition rate at low beam energies. 
The statistical accuracy
quoted here would thus scale accordingly.}
 of 15 fb$^{-1}$ for the low energy running of the
Compton collider. The polarizations of the incident electron beams and the
laser photons are chosen such that the product of the helicities $\lambda_e
\lambda_{\gamma} = -1$. 
This ensures high monochromaticity and polarization of the photon beams 
\cite{t,tp}.
Within this scenario a typical resolution of the Higgs mass is about 10~GeV, so
that for comparison with
the background process $BG \equiv \gamma \gamma \longrightarrow q \overline{q}$
one can use \cite{bbc}
$\frac{L_{\gamma \gamma}}{10\; {\rm GeV}} = \left. 
\frac{d L_{\gamma \gamma}}{dw}\right|_{m_H}$
with $\left. \frac{d L_{\gamma \gamma}}{dw} \right|_{m_H}=$0.5 fb$^{-1}$/GeV.
The number of background events is then given by
$N_{BG} = L_{\gamma \gamma} \sigma_{BG}$.
\begin{center}
\begin{figure}
\centering
\epsfig{file=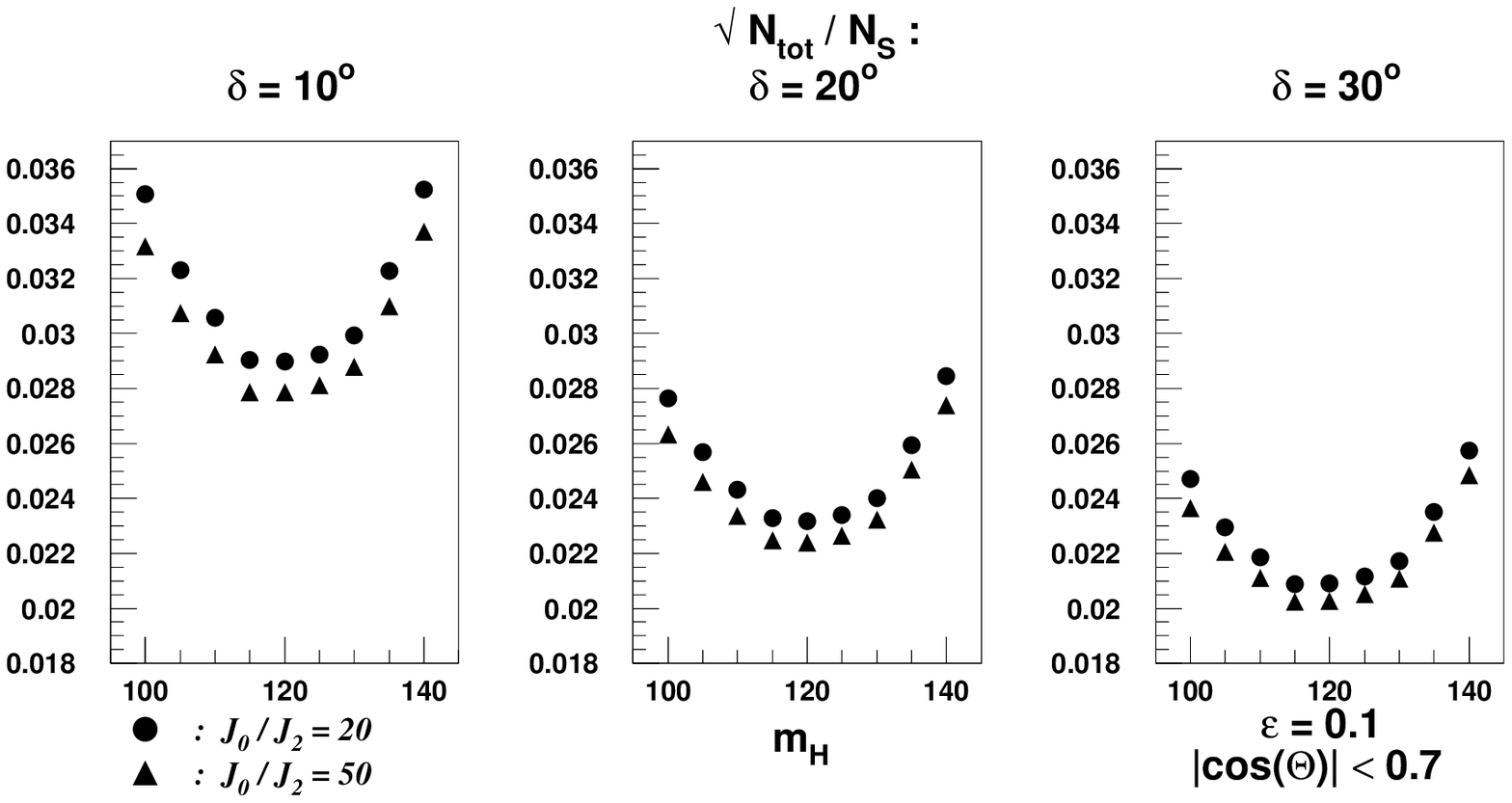,width=13cm}
\epsfig{file=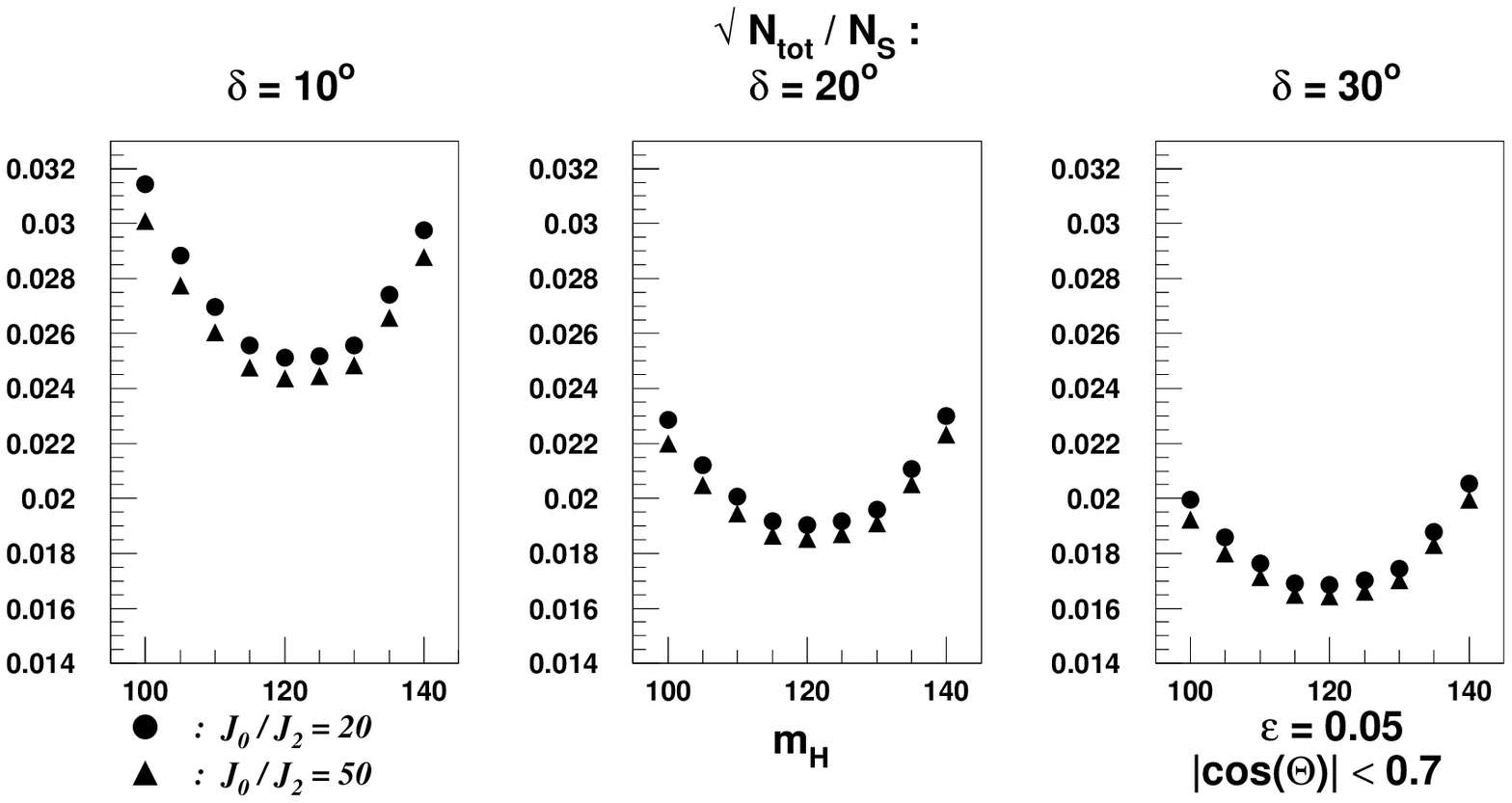,width=13cm}
\caption{The cone-angle dependence of the inverse statistical significance
of the intermediate mass Higgs signal for the displayed values of thrust
and energy cut parameters. Overall a 70\% double b-tagging efficiency and
a 1\% charm misidentification rate are assumed. For larger values of $\delta$
the number of events is enlarged, however, the theoretical uncertainty increases
.
For smaller values of $\epsilon$ higher order cutoff dependent terms might becom
e
important.}
\label{fig:ssig}
\end{figure}
\end{center}
\vspace{-1cm}
In Ref. \cite{msk} it was demonstrated that in order to achieve a large enough
data sample, a central thrust angle cut $| \cos \theta | < 0.7$ is 
advantageous and is adopted here. We also assume a (realistic) 70\%
double b-tagging efficiency. For the charm rejection rate, however,
it seems now possible to assume an even better detector performance.
The improvement comes
from assuming a better single point resolution, thinner detector modules and
moving the vertex detectors closer to the beam-line \cite{ba}.
With these results in hand we keep $| \cos \theta | < 0.7$ fixed
and furthermore assume the $\overline{c} c$ misidentification rate of 1\%
\cite{msk}. We vary
the cone angle $\delta$ between narrow ($10^o$), medium ($20^o$) and large
($30^o$) cone sizes for both $\epsilon=0.1$ and $\epsilon=0.05$.
The upper row of Fig. \ref{fig:ssig} demonstrates that for the former
choice of the energy
cutoff parameter we achieve the highest statistical accuracy for the
large $\delta=30^o$ scenario of around 2\%. We emphasize, however, that
in this case also the missing ${\mathcal O} \left( \alpha_s^2 \right)$ bremsstrahlung
corrections could become important.
The largest effect is obtained by effectively suppressing the background
radiative events with the smaller energy cutoff of $\epsilon=0.05$ outside
the cone (the inside is of course independent of $\epsilon$). Here the
lower row of Fig.
\ref{fig:ssig} demonstrates that the statistical accuracy of the Higgs
boson with $m_H < 130$ GeV can be below the 2\% level after collecting one
year of data. We should mention again that for this choice of $\epsilon$
we might have slightly enhanced the higher order (uncanceled) cutoff dependence.
The dependence on the photon-photon polarization degree is visible but not
crucial. We also conclude that
the good charm misidentification rate is important
for $\sqrt{N_{tot}}/N_S$.
Together with the expected uncertainty of 1\% from the $e^+e^-$ mode determination
of BR$(H \longrightarrow \overline{b} b)$,
we conclude that a measurement of the partial width
$\Gamma (H \longrightarrow \gamma \gamma)$ of 2\% precision
level\footnote{We assume uncorrelated error progression and negligible
systematic errors.}
is feasible for the MSSM mass range from a purely statistical point of view.
With the aforementioned possible luminosity increase by a factor of 15 
\cite{tgg}, this number could come down by a factor of two and would be
dominated by the error on BR$(H \longrightarrow \overline{b} b)$.
This level of accuracy could significantly enhance the kinematical reach
of the MSSM parameter space
in the large pseudoscalar mass limit and thus open up a window for
physics beyond the Standard Model.
In summary, using realistic and optimized machine and detector design parameters, we
conclude that the Compton collider option at a future linear collider
can considerably extend our ability to discriminate between the SM and
MSSM or 2HDM scenarios.
\section*{Acknowledgments}
I would like to thank my collaborators W.J.~Stirling and V.A.~Khoze for their contributions to 
the results presented here. In addition I am grateful for interesting discussions with 
G.~Jikia and V.I.~Telnov.

\end{document}